# Magnitude-image based data-consistent deep learning method for MRI super resolution

Ziyan Lin
*High School*
*Shanghai Starriver Bilingual School*
Shanghai, China
lily.ziyan.lin@gmail.com

Zihao Chen
*Department of Bioengineering*
*University of California, Los Angeles*
Los Angeles, USA
zihao19@ucla.edu

*Abstract*— **Magnetic Resonance Imaging (MRI) is important in clinic to produce high resolution images for diagnosis, but its acquisition time is long for high resolution images. Deep learning based MRI super resolution methods can reduce scan time without complicated sequence programming, but may create additional artifacts due to the discrepancy between training data and testing data. Data consistency layer can improve the deep learning results but needs raw k-space data. In this work, we propose a magnitude-image based data consistency deep learning MRI super resolution method to improve super resolution images' quality without raw k-space data. Our experiments show that the proposed method can improve NRMSE and SSIM of super resolution images compared to the same Convolutional Neural Network (CNN) block without data consistency module.**

## I. Introduction

Medical Imaging Devices are often used in clinical scenarios to help with diagnostic process. Magnetic Resonance Imaging (MRI) is considered one of the most important modalities for its high resolution and good contrast in soft tissues; however, it requires long acquisition time for scanning high resolution images (especially for 3D images). One conventional method to reduce the acquisition time is using the compressed sensing method [1], which is to utilize random undersampling pattern to reduce the acquisition time. However, the method needs to change acquisition scheme and parameter fine-tuning in reconstruction, which makes it difficult to be applied in many clinical settings.

Recently, deep-learning based MR super resolution methods have been proven to be good ways of reducing scan time without complicated sequence programming [2-5]. After quick low-resolution (LR) image acquisition, the neural network can efficiently output the corresponding super-resolution (SR) image. However, the performance of purely data-driven neural network is solely based on its training data, and may create additional artifacts in the testing due to the discrepancy between training data and testing data.

In the realm of deep learning MR reconstruction, data consistency (DC) layer [6-8], which incorporates the raw k-space data into the neural network, has been applied to improve the deep learning reconstruction's accuracy over purely data-driven models. However, to the best of our knowledge, there is no previous work on using DC layer in deep learning MR super resolution. One potential reason for that is most previous MR super resolution works only have magnitude images for training and lack raw k-space due to the dataset availability.

In this work, we propose a magnitude-image based DC deep learning MR super resolution method and show that it is feasible and beneficial to apply DC layer without raw k-space in MR super resolution. When the phase variation of a complex MR image is small, there is not much difference between magnitude image and complex image. Even if there is difference, the network can learn to compensate for that part in supervised learning. Therefore, in the case of small image phase variation, we propose that we can use the k-space of magnitude LR image to approximate the raw k-space and apply the DC layer without the true raw k-space.

For comparison, we have also trained a SR network that has the same structure as the proposed method but without the proposed magnitude-image based DC layer. The Normalized Root Mean Square Error (NRMSE) and Structural Similarity (SSIM) were used to evaluate the SR imaging qualities. The proposed method is shown to improve the SR image quality and reduce the artifacts over the SR model without DC layer.

## II. Theory

### A. Problem Formulation

The MR super resolution problem can be treated as a special case of undersampling MR reconstruction, in which

the sampling mask is central k-space lines. The acquisition process can be written as:

$$b = MFx + \epsilon \qquad (1)$$

Here $x \in \mathbb{C}^N$ is the vectorized original high-resolution (HR) MR image ($N = N_x \times N_y$), $F$ is the Fourier Transform operator, $M \in \mathbb{R}^{M \times N}$ ($M < N$) is a sampling mask matrix composed of 1 and 0, $b$ is the vectorized acquired k-space data, and $\epsilon$ is the random noise during the acquisition.

Conventionally, the regularized minimization is used to solve $x$ from $b$:

$$x = \arg\min_x \|MFx - b\|_2^2 + \lambda R(x) \qquad (2)$$

Here $R$ is a regularization term to help solve the ill-posed problem, and $\lambda$ is the regularization factor. In compressed sensing [1], $R$ typically involves L1 norm in the sparsifying domain of $x$.

*B. Data Consistency Layer*

In some recent deep learning reconstruction works, data consistency (DC) layer [6, 7] was proposed to improve the reconstruction's performance. In DC layer, the convolutional neural network (CNN) prediction image $x_{cnn}$ is combined with the data fidelity term in (2), and the new minimization problem becomes:

$$x_{dc} = \arg\min_x \|MFx - b\|_2^2 + \lambda \|x - x_{cnn}\|_2^2 \qquad (3)$$

In the case of Cartesian undersampling, (3) has a closed form solution:

$$s_{dc}(i) = \begin{cases} s_{cnn}(i) & if\ i \notin \Omega \\ \frac{\lambda s_{cnn}(i) + s_0(i)}{1 + \lambda} & if\ i \in \Omega \end{cases} \qquad (4)$$

Here $s_{dc} = Fx_{dc}$, $s_{cnn} = Fx_{cnn}$, $s_0 \in \mathbb{C}^N$ is the zero-padded vectorized acquired k-space. $\Omega$ is the set of the acquired k-space points.

Therefore, (4) can be used to construct DC layer in the deep learning MR reconstruction when the acquired k-space $b$ and the sampling mask $M$ are known.

*C. DC Layer for Magnitude MR Super Resolution*

The raw k-space data are usually unavailable in clinical settings because the scanner usually exports magnitude DICOM images to radiologists. To be a clinical practical MR super resolution method, it will be the best to only use magnitude low resolution (LR) images as inputs, without requiring raw k-space data.

The magnitude LR image can be expressed as:

$$x_{LR}^M = |F^{-1} MF x_{HR}^C| \qquad (5)$$

Here $x_{HR}^C$ represents the complex HR image (the superscript $C$ means complex while $M$ means magnitude).

In general, $Fx_{LR}^M = F|F^{-1}MFx_{HR}^C| \neq MFx_{HR}^C$, which means we cannot treat the Fourier Transform of the magnitude LR image as the raw k-space data $b$, and the DC layer in (3) cannot be formulated.

However, if $x_{HR}^C$ has a very small phase variation and can be approximated to a real-value image (i.e., $x_{HR}^C \approx |x_{HR}^C|$), then $Fx_{LR}^M = F|F^{-1}MFx_{HR}^C| \approx MFx_{HR}^C$ when $M$ is a symmetric sampling mask to the k-space center. Next, we can replace $b$ with $Fx_{LR}^M$ in (3):

$$x_{dc} = \arg\min_x \|MFx - Fx_{LR}^M\|_2^2 + \lambda \|x - x_{cnn}\|_2^2 \qquad (6)$$

The solution of (6) has the same form as (4), but the $s_0$ is no longer the acquired k-space but the k-space vectorized from $Fx_{LR}^M$. Therefore, we can have a magnitude-image based DC layer for MR super resolution.

It is known that the phase variation in MRI often comes from B0 field inhomogeneity and flows. Therefore, $x_{HR}^C \approx |x_{HR}^C|$ can be established when the scanner has good shimming and there is no major flow in the imaging slice.

## III. EXPERIMENT

*A. Datasets and preprocessing*

We use NYU fastMRI dataset in this work [9]. A total of 720 2D knee image slices were used in this work. We divided the dataset into training, validation and testing sets with the ratio of 8:1:1 (580 slices for training, 70 slices for validation, and 70 slices for testing).

The original k-space in the fastMRI dataset is regarded as the k-space of HR reference images. To generate the k-space of LR inputs, we crop the center of HR k-space in the phase encoding direction (Fig. 1), and the undersampling factor is 4. Then both HR k-space and LR k-space are converted to magnitude images by inverse Fourier Transform and taking absolute value (Fig. 1). Both LR input and HR reference images are normalized by their 95 percentiles before training and testing.

We have calculated the phase variation in our dataset to make sure that the small phase variation condition in session II-C holds for our dataset. The average phase variation in the whole dataset is 36.8°, which is considered as a small phase variation.

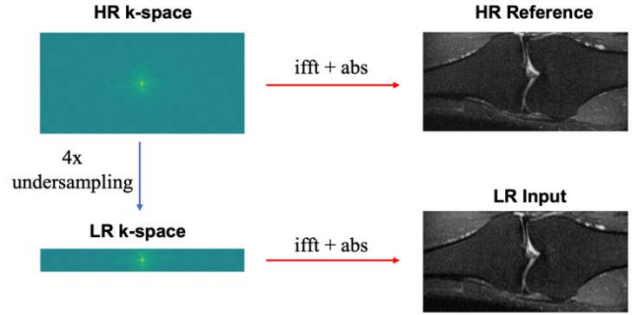

Fig. 1. The imaging preprocessing method to generate HR reference and LR input images

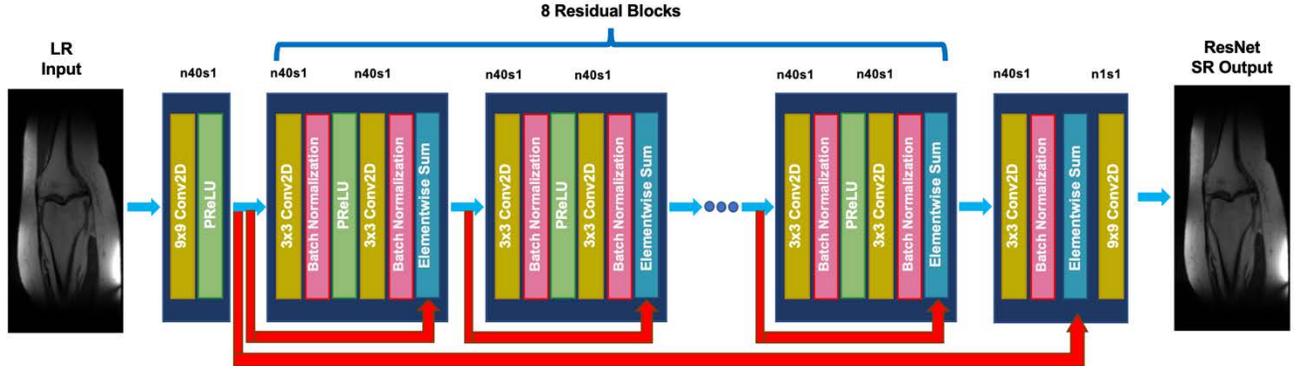

Fig. 2. The network architecture of the super resolution ResNet. (n) denotes to number of filters and (s) denotes to stride.

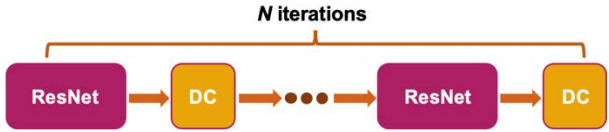

Fig. 3. The unrolled network structure in the proposed method.

## B. Implementation Details

The proposed method's network consists of two parts: the ResNet [10] super resolution module (Fig. 2) and the magnitude-image based DC layer that is described in session II-C. The $\lambda$ in (4) is set to be a trainable parameter in our DC layer. An unrolled network structure with several iterations is used in the proposed method to combine ResNet module and DC layer (Fig. 3). For all the iterations in the unrolled network, the ResNet modules share the same parameter values.

To evaluate the benefit of the proposed DC layer, we trained another network that only has the ResNet in Fig. 2 (no DC layer) for comparison.

Experiments are conducted through TensorFlow with one NVIDIA GeForce RTX 3080 GPU. The loss function is the mean absolute value (MAE) between SR output images and HR references images. The optimizer is Adam, and learning rate is set to 2e-4. We trained the networks for 35 epochs respectively for both the proposed network and the ResNet without DC.

For the proposed unrolled network, the number of iterations $N$ in Fig. 3 can be set to different values to assess the influence of number of iterations on the final image quality. In our experiment, we have trained and evaluated for $N = 1, 2, 3, 4$. For each $N$ value, the whole unrolled network was retrained for 35 epochs.

## C. Evaluation

Normalized Root Mean Square Error (NRMSE) and Structural Similarity (SSIM) were used to evaluate the SR imaging quality of the proposed unrolled model and the ResNet-only model. The original HR images in Fig. 1 were the reference images when computing NRMSE and SSIM. Paired t-test was used to determine the statistical significance of different values.

To visually evaluate the imaging quality of different SR models, we will show one example slice for different SR images and the corresponding LR and HR images.

## IV. RESULTS

The NRMSE and SSIM of different SR models among the testing set are shown in Table I. It can be observed that ResNet without DC improves the imaging quality from LR input, and the unrolled model improves the imaging quality from ResNet w/o DC. The unrolled model's performance is affected by the number of iterations ($N$), and $N = 1$ or $2$ shows the best results in our experiment. For the unrolled model with 1 or 2 iterations, the NRMSE and SSIM are statistically significantly better than ResNet w/o DC (p<0.001).

TABLE I. QUANTITATIVE IMAGING METRICS OF DIFFERENT SR METHODS AMONG THE TESTING SET. THE NUMBERS IN THE BRACKETS ARE STANDARD DEVIATIONS. THE BEST VALUES ARE BOLD.

|  | NRMSE | SSIM |
|---|---|---|
| **LR Input** | 0.218 (0.157) | 0.682 (0.210) |
| **ResNet w/o DC** | 0.204 (0.116) | 0.698 (0.217) |
| **Unrolled model (1 iteration)** | **0.189 (0.123)** | 0.707 (0.220) |
| **Unrolled model (2 iterations)** | 0.190 (0.122) | **0.709 (0.221)** |
| **Unrolled model (3 iterations)** | 0.204 (0.120) | 0.705 (0.220) |
| **Unrolled model (4 iterations)** | 0.194 (0.122) | 0.706 (0.217) |

Fig. 4 shows an example image slice of SR images from different models as well as the corresponding LR and HR images. It can be clearly observed that ResNet w/o DC may produce some artifacts in the SR output, but the proposed unrolled model can remove such artifacts, improving the SR output's fidelity and imaging quality.

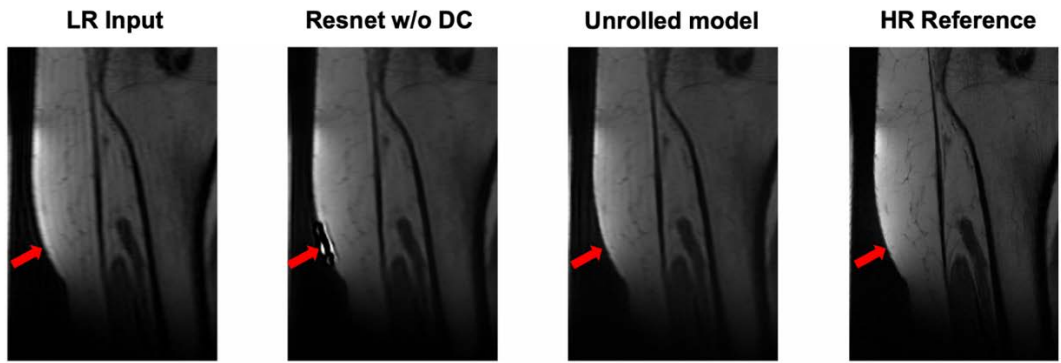

Fig. 4. An example image slice of SR images from different models as well as the corresponding LR and HR images. In the unrolled model, the iteration number was set to 2 in this example.

## V. Discussion

In this work, we have shown that the proposed magnitude-image based DC layer can improve the imaging quality of deep learning MR super resolution even without the information of raw k-space data. The magnitude-image based DC layer has significantly enlarged DC layer's scope of application because most clinical MR images don't come with raw k-space data. And the magnitude-image based DC layer can be applied to other fields like MR reconstruction.

The focus of the work is to show the improvement of magnitude-image based DC layer in MR super resolution. We haven't fine-tuned the CNN structure, and we only use small network and small dataset for simplicity. But our DC layer can be implemented with any advanced CNN blocks and improve their performance.

According to the results in Table I, it is good to use small iteration numbers like 1 or 2 when applying the proposed method. The results in Table I shows that the proposed unrolled model's performance does not monotonically improve with the increasing of iteration numbers, which is different from the results of the previous DC layer works [6]. This difference may be explained by the difference between the proposed magnitude-image based DC layer and the previous raw k-space based DC layer. Our magnitude-image based DC layer is not fully accurate in the data fidelity term as was shown in session II-C, so the reconstruction error may accumulate after the increasing of iteration numbers. However, this issue does not hurt the benefits of the proposed method in the case of no available raw k-space, because previous DC layers cannot even operate when the raw k-space data are absent.

For future work, we will compare the proposed magnitude DC layer vs. complex DC layer and include more recent SR networks for comparison. Meanwhile, we will research the relationship between the performance of magnitude DC layer and phase variation strength.

## VI. Conclusion

In this work, we developed a magnitude-image based DC layer to improve deep learning MR super resolution without raw k-space data. Our experiments show that the proposed method can improve NRMSE and SSIM of SR images compared to the same CNN block without DC layer.

Previous DC network requires raw k-space data, which are not included in many clinical MR scans and public MR datasets. The proposed method may open a new door for the application of DC layer in MR super resolution and reconstruction since the available dataset are significantly enlarged with magnitude images.